\documentclass[11pt,a4paper]{article}
\usepackage{jheppub}
\usepackage{xfrac,float,cancel,booktabs}
\usepackage{enumitem}

\DeclareFontFamily{U}{BOONDOX-cal}{\skewchar\font=45 }
\DeclareFontShape{U}{BOONDOX-cal}{m}{n}{
	<-> s*[1.05] BOONDOX-r-cal}{}
\DeclareFontShape{U}{BOONDOX-cal}{b}{n}{
	<-> s*[1.05] BOONDOX-b-cal}{}
\DeclareMathAlphabet{\maths}{U}{BOONDOX-cal}{m}{n}
\SetMathAlphabet{\maths}{bold}{U}{BOONDOX-cal}{b}{n}
\DeclareMathAlphabet{\mathbs}{U}{BOONDOX-cal}{b}{n}

\title{A Novel Solution to the Gravitino Problem}

\author[a,b]{Yu-Cheng Qiu}
\author[a,c]{and S.-H. Henry Tye}

\affiliation[a]{Jockey Club Institute for Advanced Study,\\
Hong Kong University of Science and Technology, Hong Kong S.A.R., China}

\affiliation[b]{Tsung-Dao Lee Institute,\\
Shanghai Jiao Tong University, 520 Shengrong Road, Shanghai, 201210, China}

\affiliation[c]{Department of Physics,\\
Cornell University, Ithaca, NY 14853, USA}

\emailAdd{yqiuai@connect.ust.hk}
\emailAdd{sht5@cornell.edu}

\abstract{In a general phenomenological model with local supersymmetry, the amount of massive gravitinos produced in early universe tends to violate the known dark matter density bound by many orders of magnitude. In the brane world scenario in Type IIB string theory, we propose a novel way to evade this problem. There, the standard model of strong and electroweak interactions live inside the anti-${\rm D3}$-branes ($\overline{\rm D3}$-branes) that span the 3 large spatial dimensions.  Here, the ``potential" Goldstino to be absorbed by the gravitino (to become massive) is the fermion component of the open string nilpotent superfield $X$ (i.e., $X^2=0$) which is present only inside the $\overline{\rm D3}$-branes. This non-linear supergravity scenario offers 2 ways to solve the gravitino problem, with very different particle physics phenomenologies:
(1) To satisfy the necessary condition for a naturally small cosmological constant $\Lambda$, the supersymmetry breaking $\overline{\rm D3}$-branes tension is precisely cancelled by the Higgs spontaneous symmetry breaking effect, so the gravitino is ultra-light and its contribution to the dark matter density is negligible. If exist, the super-particles should have already been detected in experiments. To avoid contradiction with their non-observation, $X$ is applied to project out all the ``R-parity odd" fields.  Consequently, this non-linear supergravity model is almost identical to the standard model.
(2) As an alternative, one can have a massive gravitino (e.g., $\widehat{m}_{3/2} > 100\,{\rm GeV}$) due to the supersymmetry breaking tension of the $\overline{\rm D3}$-branes. Here, the super-particles can be heavy enough to have avoided detection so far. Since the open string Goldstino exists only inside the $\overline{\rm D3}$-branes, the gravitino is heavy only inside the $\overline{\rm D3}$-branes, but massless or ultra-light outside the $\overline{\rm D3}$-branes. This means that the gravitinos will be pushed out of the $\overline{\rm D3}$-branes to the extra dimensions in the bulk, a phenomenon analogous to the Meissner effect for the massive photons inside super-conductors but massless outside. As a result, the massive gravitinos will be depleted so the gravitino problem is absent. In this case, a fine-tuning is necessary to obtain the very small observed $\Lambda_{\rm obs}$.
}

\arxivnumber{2207.03144}

\begin{document}

\maketitle

\section{Introduction}
	
It is widely speculated that supersymmetry (SUSY) plays an essential role in the fundamental physics of particles and fields. With spontaneous supersymmetry breaking in supergravity, the production of massive gravitinos $\tilde{g}$ (the superpartner of the graviton) in the early universe has been extensively studied (e.g., see ref.~\cite{Dudas:2021njv} for a list of references). If thermally produced after the inflationary epoch, a stable gravitino contribution to the dark matter density is restricted by the known observational bounds, or ref.~\cite{Pagels:1981ke}
	\begin{equation}\label{gmassb}
m_{3/2} < 1 \,{\rm keV} \;.
\end{equation}
However, one expects the SUSY breaking scale to be much higher, especially now that not a single superpartner has been experimentally detected, i.e.,
\begin{equation}
\label{gmassX}
\widehat{m}_{3/2} \ge 100 \,{\rm GeV \,(or\, \, higher)}\;.
\end{equation}
If unstable, the gravitinos would decay~\cite{Weinberg:1982zq,Ellis:1982yb,Khlopov:1984pf}. Since they decay only through gravitational interactions, their lifetime would be long, of the order of $M_{\rm Pl}^2 / m_{3/2}^3$, where $M_{\rm Pl}$ is the reduced Planck mass. For a gravitino mass of the order of ${\rm TeV}$, this would be of the order of $10^5\,{\rm s}$, much later than the era of nucleosynthesis. At least one possible channel of decay should include either a photon, a charged lepton or a meson, each of which would be energetic enough to destroy a nucleus if it strikes one. Enough such energetic particles will be created in the decay to destroy almost all the nuclei created in the era of nucleosynthesis, in contrast with observations. Other interesting possible solutions may invoke split SUSY, R-parity violation, $B-L$ violation or low enough reheat temperature after inflation. Different non-thermal production mechanisms have also been studied; in some cases, the abundant production of massive gravitinos in non-linear supergravity may even be catastrophic~\cite{Kolb:2021xfn,Dudas:2021njv}. Since none of the proposals has been confirmed or seriously tested so far (cf.~\cite{Moroi:1993mb,Buchmuller:2019gfy}), we believe it is worthwhile to consider other possible solutions. Here, we propose a novel way.  

Consider the brane world scenario in Type IIB string theory, where we live in a stack of anti-D3 ($\overline{\rm D3}$)-branes in a flux-compactified Calabi-Yau-like orientifold. That is, all standard model (SM) particles are open string modes that exist only inside the $\overline{\rm D3}$-branes. The $\overline{\rm D3}$-branes live at the bottom of a warped throat in the 6 extra spatial dimensions (the bulk).
In contrast to ${\rm D3}$-branes, the tension of $\overline{\rm D3}$-branes breaks SUSY. The $\overline{\rm D3}$-branes introduces a nilpotent superfield $X$ ($X^2=0$), leading to a non-linear supergravity model (NSUGRA), whose fermion mode is supposedly the Goldstino to be absorbed by the gravitino as it becomes massive~\cite{Ferrara:2014kva,Kallosh:2014wsa,Kallosh:2015nia,Bergshoeff:2015jxa,Vercnocke:2016fbt,Kallosh:2016aep}. Two possible ways to solve the gravitino problem emerge in this NSUGRA model. Here are the 2 scenarios :
\begin{enumerate}[label=(\arabic*)]

	\item The starting point of the model~\cite{Sumitomo:2013vla,Li:2020rzo} is the demand that the smallness of the observed vacuum energy density (i.e., the cosmological constant) $\Lambda_{\rm obs}=10^{-120}\, M_{\rm Pl}^4$ does NOT require any extreme fine-tuning. Since $\Lambda$ is calculable in string theory and supergravity (SUGRA), we start with the above string theory motivated NSUGRA model~\cite{Li:2020rzo}. 
%, with mass bigger than $100\,{\rm GeV}$~\eqref{gmassX}. 
The usual Higgs ($H_u, H_d$) dependent potential terms in the minimal supersymmetric standard model (MSSM) or its extension (cf.~\cite{Martin:1997ns}) would require a fine-tuning to render $\Lambda$ exponentially small. To remove these undesirable terms, we use $X$ to project out these standard MSSM contributions to $\Lambda$. %, a necessary condition to avoid the fine-tuning of $\Lambda$.
%a supposedly break SUSY, where the SUSY breaking emerges from the presence of 
We are left with the superpotential $W=X(m^2-\kappa H_uH_d) + \cdots$, where, at the minimum of the potential $V$, the electroweak Higgs contribution to $\Lambda$ is precisely cancelled by that coming from the $\overline{\rm D3}$-branes tension $m^4$ that is supposed to break SUSY~\cite{Li:2020rzo}.  
Instead, the $\overline{\rm D3}$-brane tension $m^4$ drives the Higgs spontaneous symmetry breaking (SSB), which in return  screens it. The resulting $m_{3/2}$ now comes only from the Racetrack K\"ahler Uplift (RKU) model~\cite{Sumitomo:2013vla} and is exponentially small, of order 
 \begin{equation}
\label{gmassRKU}
m_{3/2} \sim 10^{-30}\, {\rm eV} \;.
\end{equation}
This is intuitively in (order-of-magnitude) agreement with the simple reasoning: a small positive $\Lambda_{\rm obs}$ implies a small SUSY breaking, or 
\begin{equation}
\label{gmassL}
m_{3/2} \sim \sqrt{{\Lambda_{\rm obs}}/{M_{\rm Pl}^2}} \sim 10^{-33} \,{\rm eV}.
\end{equation}
Since the masses of the super-particles are comparable to $m_{3/2}$, this immediately raises  the question why we have not seen any of the very light super-partners of the SM particles. This puzzle is easily solved when we use $X$ to project out all the $R$-parity odd super-particles \cite{Li:2020rzo}. As a result, this NSUGRA model has almost identical phenomenology as that of the standard model.

\item As an alternative, one may prefer to keep the super-particles in the particle spectrum. Let us start with the minimal supersymmetric model (MSSM) (or its extension) in the brane world scenario, where both the Goldstino in $X$ and the spectrum in the MSSM are open string modes. Since the gravitino mass measures the SUSY breaking scale, we choose a large gravitino mass \eqref{gmassX} so the super-particles are heavy enough to avoid laboratory detection so far. This scenario~\cite{Ferrara:2014kva,Kallosh:2014wsa,Bergshoeff:2015jxa,Vercnocke:2016fbt,Kallosh:2016aep} is within the general KKLT framework~\cite{Kachru:2003aw}. For simplicity, let us consider the superpotential $W=X M_1^2 + M_2^3$, where $M_2^3$ and the effective tension of the $\overline{\rm D3}$-branes $M_1^4$ can be functions of $H_u, H_d$. With big enough values for $M_1$ and $M_2$ (after Higgs SSB), the super-particles can be heavy enough to have avoided laboratory detection so far. 

Even with such a heavy gravitino, the usual gravitino problem is absent. The key point is that the Goldstino (an open string mode in $X$) responsible for SUSY breaking in MSSM is present only inside the $\overline{\rm D3}$-branes. In the absence of the Goldstino, the gravitinos outside the $\overline{\rm D3}$-branes do not pick up a big mass (though they may still pick up an ultra-light mass outside the $\overline{\rm D3}$-branes via another very weak SUSY breaking mechanism, e.g., that in the RKU model).
The situation is analogous to the (abelian) Higgs field (i.e., the order parameter for the Cooper pair) that is present only inside a superconductor (SC) but not outside. Analogous to the need of the Higgs field for the photon to become massive via SSB, a Goldstino is needed for the gravitino to become massive via SUSY breaking.  Analogous to the Meissner effect, where a massive photon in a SC will be pushed out of the SC to become massless outside, a massive gravitino ($\widehat{m}_{3/2} \ge 100\,{\rm GeV}$) will be pushed out of the $\overline{\rm D3}$-branes to become extremely light (e.g., with mass~\eqref{gmassRKU}) outside.  This leads to a depletion of the massive gravitinos in the universe, so the usual gravitino problem is absent. In fact, energy consideration suggests that ultra-light gravitinos are much more likely to be created than the heavy ones in early universe. Notice that, in contrast to (1), a fine-tuning is necessary to obtain $\Lambda_{\rm obs}$ in this scenario.
\end{enumerate}

The rest of the paper goes as follows. In Sec.~\ref{sec:review}, we review our model and discuss scenario (1). In Sec.~\ref{sec:gravitino_mass}, we discuss  the gravitino mass in the above two scenarios. Sec.~\ref{sec:scenario_2} discusses how a massive gravitino mass in scenario (2) can still avoid the standard gravitino problem in cosmology. Remarks can be found in Sec.~\ref{sec:remarks}. Some details are included in the two appendices.

\section{Review of the Model}\label{sec:review}
	
	Since the search within string theory for the precise description of our observed universe is very challenging, we take a step back to implement stringy elements into a phenomenological description of our universe. In this attempt, the requirement of a small vacuum energy density $\Lambda$ without fine-tuning becomes a powerful guide in the search. This leads us to a well defined NSUGRA model where the standard model (SM) of strong and electroweak interactions are built in while the dynamically determined $\Lambda$ is naturally small. %In this model, the usual gravitino problem is absent.
	
In a string theory motivated SUGRA, the effective potential is composed of a F-term and a D-term, 
\[V=V_F +V_D\;,\]
\[
V_F =e^{K} \left(K^{M \overline{N}} D_M W D_{\overline{N}} \overline{W} - {3}\frac{|W|^2}{M_{\rm Pl}^2}\right)\;,
\]
where $K$ is the dimensionless K\"ahler potential and $W$ is the superpotential with mass dimension 3. $D_M = \partial_M + \partial_M K$. At times, it is convenient to combine them to form a single function
\[
\maths{G}= K + \ln \frac{|W|^2}{M_{\rm Pl}^6} \;,
\]
which is invariant under a K\"ahler transformation
$ K \to K + f + \bar{f}$ and $W \to e^{-f} W$,
where $f$ is an arbitrary function. To be concrete, let us consider the following simple form for the K\"ahler potential,
\begin{equation}
\label{eq:Kahler}
K=-2 \log \left[(T+\overline{T})^{3/2} + \xi/2\right] + X\overline{X} /M_{\rm Pl}^2  + \cdots\;,
\end{equation}
where $T$ is the K\"ahler (volume) modulus and $\xi$ is a string theory correction that lifts the vacuum to a de-Sitter space~\cite{Balasubramanian:2004uy,Westphal:2006tn,Rummel:2011cd,deAlwis:2011dp,Sumitomo:2013vla}. The superpotential $W$ contains contributions from closed string modes $W_0$, the non-perturbative $W_{\rm NP}(T)$, the Higgs and the $X$ contributions: 
\begin{equation}
\label{eq:WRKU}
W=W_0+W_{\rm NP}(T) + \hat{\mu}H_uH_d + W(X) + \cdots\;.
\end{equation}
In the brane world scenario in Type IIB string theory, the SM particles are open string modes, while the graviton et al., are closed string modes.
In the sum over $M$ and $N$, we separate the open string (brane) modes $P, Q =1,2,\cdots$
from the closed string (bulk) mode $J, I=1,2, \cdots$,
\begin{align}
V_F &=V_F^{\rm O} + V_F^{\rm C}, \\
V_F^{\rm O}&=e^{K} K^{P \overline{Q}} D_PW D_{\overline{Q}} \overline{W}, \nonumber \\
V_F^{\rm C}&=e^{K} \left(K^{J \overline{I}} D_J W D_{\overline{I}} \overline{W} - {3}\frac{|W|^2}{M_{\rm Pl}^2}\right)\;,\nonumber
\end{align}
where the Planck suppressed term $|W|^2$ is grouped with the closed string modes and included in $V_F^{\rm C}$.
Since the Higgs contribution in the F-term $V_F^{\rm O}$ and the D-term $V_D^{\rm O}$ is of the order of $100\,{\rm GeV}$, their expectation values must be absent.
 So a naturally small $\Lambda$ (without fine-tuning) necessarily requires, without fine-tuning,
\[\left\langle \left(V_F^{\rm O}+V_D^{\rm O}\right) \right\rangle =0\;.\]
We shall first propose a model where this is explicitly achieved; then we briefly review how a small $\Lambda=\langle V_F^{\rm C} \rangle$ is statistically preferred in the string theory motivated RKU model. Note that, even when  $\langle D_H W \rangle =0$, the Higgs field will still contribute to $\Lambda$ via its presence in $W$ in $V_F^{\rm C}$.

Supersymmetry is naturally incorporated in string theory. So let us start by considering the standard minimal SUSY phenomenology (cf.~\cite{Martin:1997ns} and references in there), where the superpotential for the two Higgs doublets $H_u$ and $H_d$ is given by $W=\mu H_u H_d + \cdots$, which, together with other contributions, yields the following term in the potential $V_{\rm MSSM}=V_H^{\rm O}+V_D^{\rm O}$, 
\begin{align}
V_H^{\rm O} & = |\mu|^2 \left(|h_u|^2 + |h_d|^2\right) +  \cdots \quad \to\quad  |\mu|^2 (246 \, \rm{GeV})^2 +  \cdots \;, \label{eq:VF}\\
V_{D}^{\rm O}&= \frac{g^2+g'^2}{8}\left( \left|h_u\right|^2 - \left|h_d\right|^2 \right)^2+\frac{g^2}{2}\left|h_u^\dagger h_d\right|^2\;,
	\label{eq:VD}
\end{align}
where $g$ and $g'$ are the $SU(2)$ and $U(1)$ gauge couplings respectively. 
Here $h_u=(h_u^+, h_u^0)$ and $h_d=(h_d^0, h_d^-)$ are scalar components of the two Higgs doublet superfields $H_u$ and $H_d$ as in MSSM.
To obtain SSB, soft terms presumably from SUGRA have to be introduced. 
%\sout{ With these extra soft terms,  $h_u=(h_u^+, h_u^0)$, $h_d=(h_d^0, h_d^-)$ as in MSSM.}
With the $SU(2)$ symmetry, we choose SSB as $\langle h_u^0  \rangle$ develops a vev, followed by $\langle h_d^0  \rangle \ne 0$. The resulting $\langle V_{\rm MSSM} \rangle$ contributes to the vacuum energy density,
which is typically of order of the electroweak scale, many orders of magnitude bigger than the observed $\Lambda$. To make this contribution to $\Lambda$ negligibly small, a fine-tuning on the input parameters is necessary.
 So, to avoid such a fine-tuning, $\langle V_{\rm MSSM} \rangle$ must be absent in the potential $\langle V \rangle$ in any supersymmetric realization of string theory.  
This is achieved in the following model~\cite{Li:2020rzo}.
	
\subsection{Scenario (1): An Ultra-light Gravitino}
	\label{case1}
	
In the brane world scenario in Type IIB string theory, we live in a stack of $\overline{\rm D3}$-branes with the $3$ large observed spatial dimensions spanning our observed universe, while the remaining $6$ spatial dimensions are dynamically (flux) compactified to a Calabi-Yau-like orientifold.  In contrast to a ${\rm D3}$-brane, an $\overline{\rm D3}$-brane breaks SUSY. A positive $\Lambda_{\rm obs}$ implies that we are living in a de-Sitter space, which is possible only if SUSY is broken. The small $\Lambda_{\rm obs}$ naturally
 suggests a small SUSY breaking scale (e.g., with $m_{3/2}$~\eqref{gmassRKU}).
Since the superpartners of the SM particles have not yet been observed, we shall simply project them out from the particle spectrum. 

  It has been proposed that an $\overline{\rm D3}$-brane can lift the universe from a (supersymmetric) anti-de-Sitter space to a de-Sitter space; in this KKLT scenario~\cite{Kachru:2003aw}, all scalar fields (complex structure moduli describing the shape of the orientifold and K\"ahler modulus measures the compactified volume) as well as the dilaton (measuring the couplings) can be dynamically stabilized. Our model is built on a modified version of this framework.
	
The crucial new ingredient is that we are actually living in a stack of $\overline{\rm D3}$-branes instead of ${\rm D3}$-branes. This picture is best described in the framework of NSUGRA, where an $\overline{\rm D3}$-brane has a nilpotent superfield $X$, satisfying $X^2=0$~\cite{Rocek:1978nb,Ivanov:1978mx,Ferrara:2014kva,Kallosh:2014wsa,Bergshoeff:2015jxa}. This nilpotent condition converts a normal SUGRA to a NSUGRA, analogous to the conversion of a linear $\sigma$ model to a non-linear $\sigma$ model by implementing a constraint on the $\sigma$ fields. This nilpotent constraint removes the scalar degree of freedom in $X$, while the fermionic degree of freedom $G$ in $X$ is precisely the Goldstino that renders the gravitino massive. In global SUSY, we have~\cite{Lindstrom:1979kq,Komargodski:2009rz}
\begin{equation}
\label{X1}
	X= {GG}/{2F^X} +\sqrt{2} \theta G+\theta^2 F^X,
\end{equation}
where $G_\alpha$ is the Goldstino and $F^{X}$ is the auxiliary field. 
Writing this in terms of $\tilde{\theta} =\theta + G/\sqrt{2}F^X$, one has $X=F^X\tilde{\theta}\tilde{\theta}$, which makes explicit its nilpotent property. In terms of the SUGRA variable $\Theta$, it has the same form
\begin{equation}
\label{X2}
	X= {GG}/{2F^X} +\sqrt{2} \Theta G+\Theta^2 F^X\;.
\end{equation}
Since the expectation value $\langle G_\alpha \rangle=0$ and $\langle GG\rangle=0$, any term containing $G$ (or $\overline{G}$) will drop out in $\langle V\rangle$.  In particular, $\langle X \rangle=0$. The symmetry group of a stack of $n$ coincident $\overline{\rm D3}$-branes at a smooth point in the internal space is the non-Abelian group $U(n)=SU(n)\times U(1)/Z_n$ ($n=5$ is a reasonable choice). Now we have superfield $\widehat{X} = \sum_a X_{ij}t_a^{ij}$ in the adjoint representation of $SU(n)$ and the singlet $X$ of $U(1)$ is the nilpotent superfield that contains the Goldstino~\cite{Parameswaran:2020ukp}. 

To remove $\langle V_{H}^{\rm O} \rangle$~\eqref{eq:VF}, we impose the following constraint on the Higgs superfield: 
\[X \overline{H}_{u,d}= {\rm chiral}\;,\]
or equivalently, using the superspace covariant derivative $\overline{D}_{\dot{\alpha}}$,
$$ X \overline{D}_{\dot{\alpha}}\overline{H}_{u,d}=0\;.$$
In global SUSY~\cite{Komargodski:2009rz}, this constraint gives, for $H_u$ and $H_d$,
\begin{align}
H & = h +\sqrt{2} \theta \psi_H + \theta^2 F^H  \label{eq:6b}\\
\psi_H & =i \sigma^{\nu}\left(\frac{\overline G}{{\overline F}{}^{\overline X}}\right) \partial_{\nu} h  \nonumber \\
F^H & = -\partial_{\mu}\left(\frac{\overline G}{{\overline F}{}^{\overline{X}}}\right){\overline \sigma}^{\nu} \sigma^{\mu} \frac{\overline G}{{\overline F}{}^{\overline{X}}} \partial_{\nu} h +\frac{\overline{G}^2}{2\big({\overline F}{}^{\overline{X}}\big)^2}  \partial^2 h \;, \nonumber
\end{align}
where the corresponding expressions in NSUGRA can be found in ref.~\cite{DallAgata:2015zxp}.  Note that $F^H$ is a function of $G$ so $\langle F^H \rangle =0$. As a result, here Higgs F-term $V_{H}^{\rm O}$ contribution to $\langle V\rangle$ is reduced to zero,
\begin{equation}
\label{mu2term}
	\left\langle V_{H}^{\rm O} \right\rangle \propto  \left\langle \left|F^{H_u}\right|^2 \right\rangle + \left\langle \left|F^{H_d}\right|^2 \right\rangle = 0\;.
\end{equation}
As a byproduct, the Higgsinos are removed.  Similarly, the D-term contribution $\langle V_D^{\rm O} \rangle$ to $\langle V \rangle$ must also be removed, since the magnitude of its expectation value $\langle V_D^{\rm O} \rangle$ is also around the electroweak scale.
This is achieved by using $X$ to project out one linear combination of $H_u$ and $H_d$, 
$$X\left[\left(H_u\right)_i-\epsilon_{ij} \left(\overline{H}_d\right)^j\right]=0\;,$$
leaving behind the other linear combination identified as the single SM Higgs doublet $\phi$~\cite{Li:2020rzo}.

One can now write the superpotential in terms of functions $p(\maths{H})$ and $g(\maths{H})$, where $\maths{H}=H_u H_d$,
\begin{align}
W &= X \left[ \tilde{m}^2- p(\maths{H})\right] + g(\maths{H}) + \cdots \nonumber\\
 & \simeq X \left(\tilde{m}^2- \tilde{\kappa} H_u H_d  + \cdots \right) + \tilde{\mu} H_u H_d + \cdots\;,
\label{Wx}
\end{align}
where $\tilde{m}^4$ is the SUSY breaking $\overline{\rm D3}$-brane tension. Note that we can easily include inside $p(\maths{H})$ 
other contributions such as a chiral symmetry breaking term from QCD.
We now have~\cite{Li:2020rzo},
\begin{equation}
\left\langle V_F^{\rm O} \right\rangle = \left\langle V_H^{\rm O}\right \rangle + \left\langle V_X \right\rangle = \left\langle V_X \right\rangle = \frac{\left|\tilde{m}^2 - p(\maths{H})\right|^2}{\left(T+\overline{T}\right)^3} \quad \to \quad \left|m^2 -\kappa \phi^\dagger \phi \right|^2\;,
\label{Vx}
\end{equation}
where we keep only the leading term in $p(\maths{H})$.\footnote{If, instead of $K$~\eqref{eq:Kahler}, one may start with $K = -3\log(T+\overline{T} - X \overline{X}/M_{\rm Pl}^2 + \cdots)$~\cite{Li:2020rzo}. Then $V_X=e^K K^{X \overline{X}}D_X W D_{\overline{X}}\overline{W} = |\tilde{m}^2 - p(\maths{H})|^2/{3(T+\overline{T})^2}$. This $(T+\overline{T})/3$ difference is not important here.} The factor $(T + \overline{T})$ comes from the K\"ahler potential $K$ and $m^4$ is the reduced $\overline{\rm D3}$-brane tension. Here $m$, $\kappa$ and $\phi$ are physical quantities, not bare quantities. In putting $v=\langle \phi^0 \rangle=246$ GeV and Higgs boson mass $m_{\phi}=125$ GeV, we have $m=104.3$ GeV and $\kappa=0.36$. Note that the minimum of $V_X$ vanishes even for more general $p(\maths{H})=p(\phi)$.
% and $g(\maths{H})=g(\phi)$.
The presence of general $g(\maths{H})$ has no effect on $V_X$ since it does not couple to $X$. Although it does shift the structure of $V_H^{\rm O}$, the constraints~\eqref{eq:6b} ensure the vanishing of $\langle F^{H} \rangle$. Therefore, $\langle V_H^{\rm O} \rangle$ still vanishes for general $g(\maths{H})$. In short, because of the perfect square form of $V_X$, the SUSY breaking $\overline{\rm D3}$-brane tension $m^4$ is completely screened by the Higgs SSB, even though it is this brane tension that leads to the Higgs SSB. 

Once the electroweak Higgs contribution is precisely canceled by the SUSY breaking $\overline{\rm D3}$-brane tension in $V_X$~\eqref{Vx}, $\Lambda$ comes only from $V_F^{\rm C}=V_{\rm RKU}$ in the RKU model~\cite{Sumitomo:2013vla,Qiu:2020los} (or another equivalent model). 
The final effective potential of our model reduces to
\[
\langle V  \rangle = \langle V_H^{\rm O} \rangle+\langle V_D^{\rm O} \rangle+\langle V_X \rangle+\langle V_{\rm RKU} \rangle \quad \to \quad \langle V_X \rangle+\langle V_{\rm RKU} \rangle\;,
\]
where $\langle V_X \rangle \to 0$ at the minimum of the potential after Higgs SSB, so $ \Lambda=\langle V_{\rm RKU}\rangle$.

In the RKU model, after scanning over the ``dense discretuum" of flux values (in practice, all the parameters of the model), the properly normalized probability distribution $P(\Lambda)$ diverges at $\Lambda=0$, so a small $\Lambda$ is preferred~\cite{Sumitomo:2013vla}. If we take the median value of $P(\Lambda)$ to be the observed value 
$\Lambda_{50\%} = \Lambda_{\rm obs}$,
%\sout{$\Lambda_{\rm obs}=10^{-120}\, M_{\rm Pl}^4$,}
we find that $T + \overline{T}  \sim  10^3$~\cite{Andriolo:2019gcb,Qiu:2020los}. % and the (not reduced) brane tension $\tilde{m} \sim 10 \; {\rm TeV}$. 
With the very small $\Lambda_{\rm obs}$, one expects the final SUSY breaking scale to be very low. So the super-particles, if present in the spectrum, should be very light and should have been detected already.
 Since they have not been detected, $X$ may be used to remove them from the particle spectrum~\cite{Li:2020rzo}. In the context of string theory ($\overline{\rm D3}$-brane in an orientifold), the application of $X$ as a projection operator in this string theory scenario has been discussed in ref.~\cite{Komargodski:2009rz,Kallosh:2014wsa,Bergshoeff:2015jxa,Vercnocke:2016fbt,Kallosh:2016aep}. 

%Presumably, these $\overline{\rm D3}$-branes will disappear via tunnelling so the universe ends up in a supersymmetric vacuum. Fortunately, this tunneling/decay process will take a period much longer than the age of the universe~\cite{Kachru:2002gs}. 
 
Our scenario (1) is different from the KKLT model~\cite{Kachru:2003aw} in a number of ways : 
\begin{itemize}

\item In our model, the SM lives in a stack of $\overline{\rm D3}$-branes, a property not spelled out in the KKLT model. As we have seen, this has important consequences.

\item In the KKLT scenario, the uplift from AdS to dS comes from the $\overline{\rm D3}$-brane tension. There, a fine-tuning is necessary to obtain an exponentially small $\Lambda$. In our model, the uplift (from AdS to dS) comes from the K\"ahler (volume modulus) uplift (KU) with a stringy ($\alpha'^3$) correction~\cite{Balasubramanian:2004uy,Westphal:2006tn,Rummel:2011cd,deAlwis:2011dp}, since the SUSY-breaking $\overline{\rm D3}$-brane tension is precisely canceled by the electroweak symmetry breaking contribution. 

\item The nilpotent field $X$ allows us to remove the dangerous contributions to $\Lambda$ as well as removing any phenomenologically undesirable superpartners in the particle spectrum.

\item The stability of the K\"ahler modulus in KKLT is provided by a non-perturbative term; such a non-perturbative term naturally emerges from the gaugino condensation in a stack of $N$ D7-branes wrapping a 4-cycle~\cite{Dine:1985rz,Derendinger:1985kk,Shifman:1987ia,Berg:2004ek}, with $\mathcal{N}=1$ supersymmetric $SU(N)\times U(1)$ gauge symmetry. To have a naturally small $\Lambda$, we require at least $2$ such non-perturbative terms, where this racetrack in RKU drives $\Lambda$ to an exponentially small positive value~\cite{Sumitomo:2013vla,Qiu:2020los}. See appendix~\ref{D7A} and table~\ref{tab:d7}  for an illustration.
  
\item The requirement of a naturally small $\Lambda$ puts very tight constraints on model building. The resulting NSUGRA model ends up almost identical to the original SM, in contrast to the MSSM and its many extensions. 
  
\end{itemize}

\section{The Gravitino Mass}
\label{sec:gravitino_mass}

We are now ready to discuss the gravitino mass.
The gravitino $\Psi_{\mu}$ mass is generated by SUSY breaking and its coupling to the fermions $\psi^i$ that form the Goldstino $\eta=\maths{G}_i \psi^i$ which
is given in the Lagrangian $\maths{L}$ by
\begin{align}
e^{-1} \maths{L}= & \frac{i}{2} e^{\maths{G}/2}\left[\overline{\Psi}_{\mu}\sigma^{\mu \nu} \Psi_{\nu} + \sqrt{2} \maths{G}_i \overline{\Psi}_{\mu}\gamma^{\mu}\psi^i + \cdots \right] \nonumber \\
=&\frac{i}{2} e^{\maths{G}/2}\overline{\Psi}'_{\mu}\sigma^{\mu \nu} \Psi'_{\nu} + \cdots \;,
\end{align}
where the Goldstino $\eta$ is absorbed by the gravitino,
$$\Psi'_ {\mu}= \Psi_{\mu} -\frac{i}{3 \sqrt{2} }\gamma_{\mu} \eta -\frac{\sqrt{2}}{3} e^{\maths{G}/2} \partial_{\mu} \eta\;,$$
with mass
\begin{equation}
m_{3/2} = e^{\maths{G}/2}M_{\rm Pl}= e^{K/2}\frac{|W|}{M_{\rm Pl}^2} \;.
\end{equation}
Let us now consider scenartio (1) and scenario (2).

\subsection{Scenario (1): $W=X(m^2-\kappa H_uH_d) + \cdots$}

Scenario (1) is described in the above section, with an ultra-light gravitino  satisfying~\eqref{gmassRKU}.
This $W$ yields $V_X$ \eqref{Vx}, where $V_X \to 0$ at the minimum. We see that $\maths{G}_XG=0$ since
\[
\maths{G}_X \propto \frac{\overline{X}}{M_{\rm Pl}^2} + \frac{1}{W} \left| m^2 -\kappa  H_u H_d \right| \;\to\; 0 \;,\]
which vanishes at the minimum. This means that the SUSY breaking $m^2$ is cancelled by the Higgs contribution so the fermion $G$ in $X$ does not contribute directly to the Goldstino $\eta$.
Next consider the Higgs fields. Since the Higgsinos $\psi_H$s have been projected out, there is no $\maths{G}_H\psi_H$ contribution to $\eta$ from the Higgs sector. \footnote{With $\psi_H$ given in Eq.(\ref{eq:6b}),
$\psi_H \propto {\overline G} \partial_{\nu} h /{{\overline F}{}^{\overline X}}$ and
$\maths{G}_{H_u}\propto \overline{H}_u + \mu H_d/{W} \ne 0$,
 it seems that $G$ does appear indirectly in $\eta$. 
However, $\partial_{\nu} h $ appears in the gauge transformation of gauge fields and so may be set to zero.} 
%we see that the $G$ in $X$ does contribute (indirectly) to the Goldstino $\eta$. 

Now the gravitino mass comes entirely from the RKU model. Statistically, in scanning over all the parameters of the model, one finds the (normalized) probability distribution $P(\Lambda)$ peaks sharply at $\Lambda=0$, $P(\Lambda) \propto \Lambda^{-1+1/2N}$, where $N$ is the rank of the ${\rm D7}$-brane gauge group~\cite{Sumitomo:2013vla}. Choosing $N \sim 100$ such that the median $\Lambda_{50\%}$ equals the observed value $\Lambda_{\rm obs}$, and with typical appropriate choice of the parameters to yield the median $\Lambda_{50\%}$~\cite{Qiu:2020los}, we obtain
\begin{equation}
m_{3/2} \sim \frac{(100\, {\rm GeV})^3}{(T+\overline{T})^{3/2}M_{\rm Pl}^2} \sim 10^{-30} \,{\rm eV} \;,
\end{equation}
a result that has been quoted \eqref{gmassRKU} earlier.
For such a very small $\Lambda_{\rm obs}$, we expect intuitively a very small SUSY breaking scale and so a very light gravitino mass \eqref{gmassL}. This expectation is born out by the RKU model.

With such a small gravitino mass, the super-particles would be very light if they exist and can be easily discovered. However, none has been detected. To satisfy the non-observation of these very light super-particles, these $R$-parity odd \footnote{R-parity $P_R=(-1)^{3(B-L)+2s}$ is a function of the baryon number $B$, the lepton number $L$ and spin $s$.} particles must be removed from the particle spectrum. 
This is easy to achieve under the constraints on the quarks, leptons and gauge superfields (the Higgsinos have already been removed),
\begin{equation}
XQ_i=XL_j=XW_{\alpha} = 0 \;,\label{susyout} 
\end{equation}
so the squarks, the sleptons and the gauginos are projected out.
For example, for a quark superfield $Q=\tilde{q} +\sqrt{2}\theta q +\theta^2F_q$, one now has the quark $q$ but not the scalar-quark $\tilde{q}$, 
\begin{align}
Q=&\frac{qG}{F^X} - \frac{GG}{2(F^X)^2}F_q +\sqrt{2}\theta q +\theta^2F_q \\\nonumber
=&\sqrt{2}\left(q -\frac{F_qG}{F^X}\right)\tilde{\theta} + F_q \tilde{\theta}^2\;.
\end{align}

In summary, although this model comes from a string theory motivated NSUGRA, the resulting model has almost identical particle physics phenomenology as the SM.  

Although this is the simplest scenario, other possibilities are {\it A priori} not ruled out, since an explicit construction of the SM from string theory is still lacking. It is possible that the SUSY breaking effect is screened in $\Lambda$ but not in the supper-particle masses in the particle spectrum. If this is the case, then the supper-particle masses are of the scale 
\begin{enumerate}
\item $m \sim 100$ GeV, in which case they still have to be projected out~\eqref{susyout}; otherwise at least some of them should have already been detected at LHC; or
\item $\tilde{m}\sim 10$ TeV, in which case they are heavy enough to have avoided detection so far. In this case, the squarks, sleptons and gauginos can be present, but there is only a single Higgs doublet and without any of the Higgsinos.
\end{enumerate}

We note that the lower bound on the gravitino mass discussed in e.g., ref.~\cite{Maltoni:2015twa, Hebecker:2019csg}, does not apply to our ultra-light gravitino model in this scenario.

Consider the production of 2 gravitons ($g$) via the annihilation of a quark pair: $q \bar{q} \to g + g$, or a gluon ($\hat{g}$) pair : $\hat{g} +\hat{g} \to g + g$. Since the gravitons are closed string modes and interact with gravitational strength,
$$\sigma (q \bar{q} \to gg) \propto \frac{E^2}{M_{\rm Pl}^4}\;, $$
which is too small to be observed at LHC. Here, the intermediate state is the quark $q$. Similarly, missing energy processes like $q \bar{q} \to gg+ {\rm jet} +\cdots$  will be too small to be detected at LHC. 
Since an ultra-light gravitino ($\tilde{g}$) is entirely a closed string mode, they will be produced with gravitational strength as well. For the production of a pair of gravitinos  $q \bar{q} \to \tilde{g}  \tilde{g}$, the intermediate state is a squark $\tilde{q}$ instead of a quark $q$. If such a squark exists with mass $m \sim 100$ GeV, it should have been detected directly at LHC already. If the squark mass is $\tilde{m} \sim 10$ TeV, then $\sigma (q \bar{q} \to \tilde{g} \tilde{g}) \lesssim  \sigma (q \bar{q} \to gg)$, too small to be detected. Similarly, processes like $q \bar{q} \to \tilde{g} \tilde{g} + \cdots $ are not observable in LHC.

As proposed in ref.~\cite{Maltoni:2015twa},
 a better way to produce gravitinos without the direct gravitational suppression is to produce super-particles which then decay to ultra-light gravitinos (plus other particles). In our model, such super-particles (squarks, sleptons and gluinos with $m \sim 100$ GeV) have been projected out of the particle spectrum via the application of the nilpotent superfield $X$. So  the production of gravitinos via the production and subsequent decays of squarks, sleptons or gauginos does not happen as these super-particles are absent. For $\tilde{m} \sim 10$ TeV, the super-particles are too heavy to be produced in LHC, so this mechanism to produce ultra-light gravitinos is again absent in our model. This is in contrast to the heavy gravitino model (Scenario (2) below), where the Goldstino $G$ in $X$ is an open string mode, so the production of gravitinos can proceed via the $G$ mode, which is not gravitationally suppressed.

\subsection{Scenario (2): $W=XM_1^2 + M_2^3$}

This is our scenario (2), where the gravitino mass $\widehat{m}_{3/2}$ is heavy~\eqref{gmassX}.
To keep the existence of the super-particles in the model, we have to demand $m_{3/2}$ of order of \eqref{gmassX}~ \cite{Kallosh:2014wsa,Kallosh:2015nia,Bergshoeff:2015jxa,Vercnocke:2016fbt,Kallosh:2016aep}. Let $K$~\eqref{eq:Kahler} and 
\begin{equation}
\label{eq:W12}
W=XM_1^2 + M_2^3\;,
\end{equation} 
so $V_X \propto M_1^4$  (e.g.,  $M_1 \sim 10$ TeV) before the rescaling. With large enough $M_1$ and $M_2$, super-particles would be heavy enough to avoid experimental detection as of today. Phenomenologically, the features of this scenario will be similar to that in the MSSM. Discovery of any super-particles will be exciting.

Following the earlier discussions, we see that $M_2$ measures the gravitino mass
\begin{equation}
\widehat{m}_{3/2}  = e^{K/2}\frac{|W|}{M_{\rm Pl}^2} \simeq \frac{M_2^3}{M_{\rm Pl}^2\mathcal{V}}\;,
 \end{equation}
which is a measure of the SUSY breaking scale, while the coupling of the Goldstino $G$ to the gravitino is given by
 \begin{equation}
e^{\maths{G}/2}\maths{G}_X = e^{K/2}\frac{|W|}{M_{\rm Pl}^3}\left( \frac{\overline{X}}{M_{\rm Pl}^2} + \frac{M_1^2}{|W|}\right)  \simeq \frac{M_1^2}{M_{\rm Pl}^3\mathcal{V}}\;,
\end{equation}
 where $\mathcal{V}^2 \sim (T+\overline{T})^3 \sim 10^9$.
 We can choose $M_1$ large (i.e., of order of TeV), but a fine-tuning is necessary to produce the small $\Lambda_{\rm obs}$, 
$$\langle V  \rangle = \frac{M_1^4}{(T+\overline{T})^3} - 3\widehat{m}_{3/2}^2 M_{\rm Pl}^2 = \frac{M_1^4-3M_2^6/M_{\rm Pl}^2}{(T+\overline{T})^3}\;, $$
where we have to fine-tune $\widehat{m}_{3/2}$ (i.e., $M_2$) and/or $M_1$ so $\langle V  \rangle =\Lambda \simeq 0$. In fact, a fine-tuning is unavoidable even if we use $X$ to project out $V_H^{\rm O} +V_D^{\rm O}$. That is, the price of having the super-particles in the particle spectrum is to give up any hope of a naturally small $\Lambda$.  Scenario (2) may be considered to be within the KKLT framework.

%Here we discuss the way of SSB of the Higgs field. 
Next, let us discuss how the Higgs SSB can be incorporated into this model, i.e., how $M_1$ and $M_2$ become functions of the Higgs field. Following ref.~\cite{Dudas:2015eha}, one can write 
\begin{align*}
K & = - 2\log{\left(\mathcal{V}+\frac{\xi}{2}\right)} - \log{\left( 1 + \frac{ X + \overline{X}}{ M_{\rm Pl}} \right)} + \cdots \\
W & = W_0 + W_{\rm NP}(T) + \hat{\mu} H_u H_d + M^2 X \;,
\end{align*}
where $X^2=0$. Expand $X$-term in $K$ as
\begin{align*}
- \log{\left( 1 + \frac{ X + \overline{X}}{ M_{\rm Pl}} \right)} = - \frac{ X + \overline{X}}{M_{\rm Pl}} + \frac{ X \overline{X}}{M_{\rm Pl}^2}\;,
\end{align*}
where higher order terms vanish due to the nilpotent condition. Perform the K\"ahler transformation as
\begin{align*}
K & \to K + \frac{ X + \overline{X}}{M_{\rm Pl}} \\ 
W & \to e^{-X/M_{\rm Pl}} W  = \left(1-\frac{X}{M_{\rm Pl}}\right)W\;,
\end{align*}
which leads to
\begin{align}
K & = -2\log{\left(\mathcal{V}+ \frac{\xi}{2}\right)} + \frac{X \overline{X}}{M_{\rm Pl}^2} + \cdots \\
W & = W_0 + W_{\rm NP} (T) + \hat{\mu} H_u H_d + X \left( M^2 - \frac{\hat{\mu} H_u H_d + W_0 +W_{\rm NP}}{M_{\rm Pl}}\right)\\
&= M_2^3 + X M_1^2 \;,
\end{align}
which reproduces $W$ \eqref{eq:W12} after the Higgs SSB and the $T$ stabilization. The potential
is now given by $V_H^{\rm O}$ \eqref{eq:VF}, $V_D^{\rm O}$ \eqref{eq:VD} and $V_X$,
\begin{equation}
\label{eq:Vnew}
V=V_H^{\rm O} +V_D^{\rm O} +V_X
\end{equation}
where
\begin{equation}
V_X \propto M_1^4 \sim \left | M^2 - \frac{\hat{\mu}H_u H_d + W_0 + W_{\rm NP}}{M_{\rm Pl}} \right|^2 \;.
\end{equation}
%Together with $V_H^{\rm O} +V_D^{\rm O}$, minimizing 
Recall that there is no Higgs SSB from $V_H^{\rm O} +V_D^{\rm O}$.
Here the presence of $V_X$ in $V$~\eqref{eq:Vnew} can now drive Higgs SSB. In turn, $V_X \propto M_1^4 \ne 0$ because of the presence of $V_H^{\rm O} +V_D^{\rm O}$. 
 
A comment is in order here. This above analysis illustrates that the physics is rather sensitive to how $X$ enters in $\maths{G}=K + \ln ( |W|^2/M_{\rm Pl}^6)$. There are multiple possibilities with different physics consequences. The final effective NSUGRA clearly needs a better understanding of the underlying string theory dynamics. On the other hand, the requirement of a naturally small $\Lambda$ in NSUGRA is by itself restrictive enough that we are smoothly led to the model~\cite{Li:2020rzo} in scenario (1). 

\section{Scenario (2): A massive Gravitino with an open string Goldstino} 
\label{sec:scenario_2}

Here we consider scenario (2), where the gravitino is heavy, with mass \eqref{gmassX}.  We observe that $G$ is an open string mode that exists only inside the stack of $\overline{\rm D3}$-branes, while the graviton and its superpartner the gravitino are closed string modes that are present everywhere. \footnote{Note that we may replace $X$ by another open-string superfield $s$ (e.g., a Polonyi-like field), which exists only inside the $3$-branes.  As long as this Goldstino exists only inside the $3$-branes, the gravitino is expected to be light or massless outside the $3$-branes.}
Since the Goldstino is crucial for the gravitino to become massive, the gravitino can become massive~\eqref{gmassX} only inside the ${3}$-branes. Outside the ${3}$-branes, the ultra-light gravitino mass~\eqref{gmassL} is mostly due to another Goldstino from the closed string sector. Energy consideration suggests that ultra-light gravitinos are much more likely to be created than the heavy ones in early universe. The heavy ones will then decay to the ultra-light ones as they escape from the $\overline{\rm D3}$-branes. As a result, even the (very) low density of heavy gravitinos produced in early universe will be quickly depleted.

In this section, we shall ignore this tiny mass and treat it as massless outside the 3-branes. The picture is analogous to the Meissner effect in superconductors (SC). Simple energetic consideration suggests that the particle, be it photon or gravitino, prefers the massless state over the massive state. However, there are some important differences between the two cases. So let us first review the SC case.

\subsection{Massive Photon}

In the language of the Abelian Higgs model, a photon is massive (with mass $m_A$) inside a SC, triggered by SSB, while it stays massless outside the SC, as there is no Higgs field and so no SSB outside the SC. 
Choosing the gauge $\partial^{\nu}A_{\nu}=0$ in the Abelian Higgs model, we have
\begin{equation}
\label{KGA}
\left(\partial_t^2 -\nabla^2 + m^2\right)A_{\mu}=0 \;.
\end{equation}
Let us consider $x$-dependent mass $m(x)$, where $m(x)=m_A$ for $x>0$ (i.e., inside SC) and $m(x)=0$ for $x\leq 0$ (outside). 
	
In a (time-independent) steady state, a solution for $A_{\mu}$ takes the form
\begin{equation}
\label{SCMeissner}
A_{\mu}(x) \simeq A_{\mu}(x=0)e^{-m_Ax} \quad \quad {\rm for} \quad x>0\;,
\end{equation}
where $A_{\mu}$ takes some non-zero value at $x \leq 0$. 
Choosing $A_1(x)=0$ to satisfy charge conservation $\partial^\mu(m^2(x)A_{\mu})=0$, and with $B_3 \simeq \partial_1A_2(x)$, one obtains, for the magnetic field ${\bf B}_{\bot}$ perpendicular to $\hat{x}$, ${\bf B}_{\bot}(x)= {\bf B}_{\bot}(x\lesssim 0)e^{-m_Ax}$.
This implies that the magnetic field inside a SC is suppressed, which is the familiar Meissner effect. That is, massive photons inside the SC will be expelled, as observed in experiments.
Dimensionally the lifetime of a massive photon is $t_A=1/m_A$. Sending a (massless) photon towards a SC will be reflected or absorbed (or tunnel through if the SC is a thin slab).

\subsection{Massive Gravitino}
\label{sec:mG}

Analogous to the massive photon case, based on energetic arguments, we expect gravitinos to stay outside the $\overline{\rm D3}$-branes, where it is (almost) massless. However, there are some differences between the two cases.	
Let us start with the gravitino equation
\begin{equation}
\label{Gravitinoeq}
\left(\gamma^{[\mu \nu \rho]} \partial_{\nu}- \widehat{m}_{3/2} \sigma^{\mu \rho} \right)\Psi_{\rho}=0\;,
\end{equation}
where the gravitino field $\Psi_{\rho}$ is a Majorana vector-spinor.
Applying the Dirac operator $\left(\gamma_{[\alpha \beta \mu]} \partial^{\beta} - \widehat{m}_{3/2} \sigma_{\alpha \mu}\right)$ on the gravitino equation~\eqref{Gravitinoeq} yields  
\begin{equation}
\label{Gravitinoeq1}
\left(\partial_t^2 -\nabla^2 +\widehat{m}_{3/2}^2(x^{\rho})  \right)\Psi_{\alpha}=0\;,
\end{equation}
where the $\partial_{\beta} \widehat{m}(x^{\rho})$-term is proportional to $\gamma^{\nu}\Psi_{\nu}$ and so absent due to the primary constraint $\gamma^{\nu}\Psi_{\nu}=0$. As expected, the gravitino field obeys the same Klein-Gordon-like equation \eqref{KGA} as the gauge field in the Abelian Higgs model.

Let us choose the coordinate $y$ to be one of the directions perpendicular to the $3$ large spatial directions. So the $y$-dependent mass $m(y)$ is $m(y)=\widehat{m}_{3/2}$ for $y>0$ (inside the $\overline{\rm D3}$-branes) while $m(y)=0$ for $y<0$ (outside the $\overline{\rm D3}$-branes). Following the massive photon case \eqref{SCMeissner}, a time-independent solution yields
\begin{equation}
\label{Gravitinoeq3}
\Psi_{\rho}(y)=\Psi_{\rho} (y \lesssim 0)e^{-\widehat{m}_{3/2}y} + \cdots \;  \, ?
\end{equation}
%which resembles~\eqref{SCMeissner} for the massive photon case; 
%where $\overline{m}_{3/2}$ is the mass inside the stack of $\overline{\rm D3}$-branes.
That is, a massive gravitino will be expelled from the $\overline{\rm D3}$-branes and become (almost) massless outside the $\overline{\rm D3}$-branes. Dimensionally, the time scale it takes to expell such a massive gravitino is $1/\widehat{m}_{3/2}$. For any mass bigger than $1\,{\rm eV}$, it is fast enough with respect to the cosmological time.

However, there are some important differences between the massive photon case and the massive gravitino case:

\begin{itemize}

\item The compactification in the $x^4$--$x^9$ directions of the orientifold renders the bulk size to be finite, so the momentum $k_j$ along these directions are discrete. Because the $\overline{\rm D3}$-branes live at the bottom of a warped throat, the Kaluza-Klein (KK) excitation (i.e. $k_j \ne 0$) levels are dictated by the size of the bottom in the $x^4$--$x^8$ directions. The width of the stack of $\overline{\rm D3}$-branes in the $y$ direction is finite,  so the $\Psi_{\rho}(y)=\Psi_{\rho} (y \lesssim 0)e^{+\widehat{m}_{3/2}y}$ solution may not be ignored in~\eqref{Gravitinoeq3}, in contrast to the massive photon case~\eqref{SCMeissner}. In appendix~\ref{MGF}, we consider the simple case where the warp factor is absent. The time-dependent solution for $\Psi_{\rho}$ in one spatial dimension yields some basic features we summarize here. For $0 \le y < l$, $m = \widehat{m}$ while $m=0$ for $l\le y <L$, where $L$ is the finite size of the bottom. Here the mass is given by the eigenvalue $\omega$ (see appendix~\ref{MGF}),
\begin{equation}
\omega^2 \simeq \left(\frac{2\pi n}{L}\right)^2 + \widehat{m}^2\left(\frac{l}{L}\right)\;.
\end{equation}
So for the ground state wavefunction, $\omega_0 \simeq \widehat{m} \sqrt{l/L}$. We see that the ground state energy (i.e., the gravitino mass here) carries a fraction of the gravitino mass $\widehat{m}_{3/2}$. That is, it is not $m=0$ outside the $\overline{\rm D3}$-branes.

\item Fortunately, what the warp geometry complicates the warp geometry saves.
Consider for example the Klebanov-Strassler throat~\cite{Klebanov:2000hb}. It is a deformed conifold whose bottom takes the topological form of $S^3 \times S^2$, where the $S^3$ attains a size of $\bar{L}_3>0$. Since SUSY is broken, it may also be resolved where the $S^2$ has a size of $\bar{L}_2>0$. When the $\overline{\rm D3}$-branes sit at the bottom of such a throat at $r=r_b$, the mouth of the throat (at $r=Z$) is glued to the bulk; so crudely, we have the warp factor $e^{-A(r)} \sim {r}/{Z}$. Comparing the string scale $M_S$ to the SUSY breaking or the electroweak scale, we consider a range of values for the warp factor (with $m \sim 100$ GeV and $\tilde{m} \sim 10$ TeV)
\[
e^{-A(r_b)} \simeq \frac{r_b}{Z} \sim \frac{m}{M_S} \text{--} \frac{\tilde{m}}{M_S} \quad\to \quad 10^{-14} \text{--}  10^{-11}\;.
\]
Consider the simple warp metric given by
\begin{equation}
ds^2= G_{MN} dx^M dx^N = e^{-2A(r)} g_{\mu\nu} dx^\mu dx^\nu + e^{2A(r)}\left( dr dr + g_{ij} dy^i dy^j \right)\;,
\end{equation}
where $G_{MN}$ is block diagonal,
\begin{equation}
G_{MN}= {\rm diag} \left(e^{-2A(r)} g_{\mu\nu}, e^{2A(r)} , e^{2A(r)} g_{ij} \right)\;.
 \end{equation}
To simplify the discussion, we shall consider a scalar mode $\Phi$ with the flat 4-dim spacetime metric $g_{\mu \nu}=\eta_{\mu \nu}$ and flat 5-dim metric $g_{ij}=\delta_{ij}$ at the bottom of the throat at $r=r_b$. 
The equation of motion $G^{MN}\partial_M \partial_N \Phi - M^2\Phi=0$ for $\Phi(x^\mu,y^i,r)$ in the warp geometry is reduced to,
\begin{equation}
\left[\eta^{\mu\nu}\partial_\mu \partial_\nu - M^2e^{-2A(r_b)}  + e^{-4A(r_b)} \partial_i^2 \right] \Phi(x^\mu,y^i,r_b)=0\;,
\end{equation}
Sitting at the bottom of the throat (at fixed $r_b$), let us choose $e^{-A(r_b)} \sim 10^{-14}$, so $m \sim M_S e^{-A(r_b)} \sim 100$ GeV; for the compactified $y^i$-directions, one would obtain a series of KK states, which obey 
\begin{equation}
\left[\eta^{\mu\nu} \partial_\mu \partial_\nu - m^2 -\sum_i e^{-4A(r_b)}\left(\frac{n_i}{L_i}\right)^2 \right]\Phi_{\{n_i\}}(x^\mu,r_b)=0\;,
\end{equation}
where $n_i$ ($i=1,2, . . .,5$) labels the wave number associated to $y^i$-direction, and $L_i$ is the size of the $y^i$-dim. Here $m=\widehat{m}_{3/2}$ inside the $\overline{\rm D3}$-branes but $m=0$ outside the $\overline{\rm D3}$-branes but still inside the bottom of the throat.

In the presence of the warp factor, the effective size of the bottom of the throat becomes $\bar{L}_i = e^{2A(r_b)}L_i \sim e^{2A(r_b)}r_b$, where we expect $L_i \simeq r_b$. 
Although $n_i$ indicates a discrete spectrum, the effective spacing of the KK levels are warped to extremely small values by the warp factor.  That is,
$$k_i= \frac{2 \pi n_i}{\bar{L}_i} = \frac{2 \pi n_i}{e^{2A(r_b)}L_i} \sim \frac{n_i}{e^{2A(r_b)}r_b} \sim \frac{n_i}{10^{+14}Z}\;.$$ 
So the momentum $k_i$ in the $y_i$ directions behave as if they are continuous, i.e., as if there is no finite size constraint. This reduces the situation back to that analogous to  the massive photon case in the Abelian Higgs model in SC. Instead of \eqref{Gravitinoeq3}, we now have
\begin{equation}
\Psi_{\rho}(y)=\Psi_{\rho,0} \exp \left({-ye^{2A(r_b)} \widehat{m}_{3/2}}\right) \simeq \Psi_{\rho, 0}\exp\left({-y M_S e^{A(r_b)}}\right)\;,
\end{equation} 
where $\Psi_{\rho,0}$ is a typical amplitude outside the $\overline{\rm D3}$-branes and $y$ measures the penetration into the $\overline{\rm D3}$-branes. This huge $e^{2A(r_b)} \sim 10^{22}$--$10^{28}$ factor in the exponent means the gravitino penetration into the branes is very severely suppressed, so the throat bottom size effect can be ignored.

\end{itemize}

The following picture emerges. Energy consideration suggests that ultra-light gravitinos are much more likely to be created than the heavy ones in early universe. So the production of the heavy ones will be suppressed. Even then, the heavy ones will eventually decay to the ultra-light ones as they escape from the $\overline{\rm D3}$-branes. Some possible decay modes are:
$\tilde{g}_{\rm heavy} \to \tilde{g}_{\rm light} + \gamma$ or any other SM particles. If energetically allowed, we may have $\tilde{g}_{\rm heavy} \to g + \tilde{\gamma}$ or any other super-particles.
Its decay lifetime is of order of $M_{\rm Pl}^2/\alpha m_{3/2}^3$. Because of its very low density, we assume such decays of the heavy gravitinos will not disrupt nucleosynthesis. 
 
 A mode with energy $\omega \ge \widehat{m}_{3/2}$ can move in and out of the $\overline{\rm D3}$-branes easily, while a mode with $\omega < \widehat{m}_{3/2}$ outside the $\overline{\rm D3}$-branes does not enter the $\overline{\rm D3}$-branes.
For a massive gravitino inside the $\overline{\rm D3}$-branes, the expansion of the universe red-shifts its momentum $p_i$ in the 3 large directions; so $p_i \to 0$. Together with scatterings, the momenta in the $y_i$ directions $k_i \to 0$ (KK decays), so its energy inside the $\overline{\rm D3}$-branes $\omega \to \omega_0 =\widehat{m}_{3/2}$ as it becomes non-relativistic.
%Sitting inside the $\overline{\rm D3}$-branes is unstable. 
Once it moves out of the $\overline{\rm D3}$-branes, its effective mass \eqref{gmassRKU} is tiny but still with energy $\omega_0$ while picking up some momenta $p_i$ and $k_j$ through scatterings. A tiny redshift (and/or further decay) will render its energy $\omega <\widehat{m}_{3/2}$ and stop it from re-entering into the $\overline{\rm D3}$-branes. Futhermore, there are $5\times 2$ directions for a massive stationary gravitino to leave the $\overline{\rm D3}$-branes, but it must point in a very specific direction to hit the $\overline{\rm D3}$-branes from outside. Any delay will lead to $\omega < \widehat{m}_{3/2}$, so it will be stuck outside $\overline{\rm D3}$-branes with an exponentially small mass. As the universe expands, massive gravitinos produced inside the $\overline{\rm D3}$-branes will move out and become (almost) massless, while the ultra-light gravitinos produced outside the $\overline{\rm D3}$-branes will simply stay outside.

\subsection{Comments}

To end this section, we like to make some comments.
\begin{itemize}
\item Even though a gravitino $\tilde{g}$ outside the $\overline{\rm D3}$-branes is extremely light, the gravitinos we can detect will always be heavy, with mass \eqref{gmassX}, since we (the SM) live inside the $\overline{\rm D3}$-branes where mass measurements are carried out.  They can be produced via scatterings of the standard model particles such as quarks $q$ and gluons, e.g., the process $qq \to \tilde{q}\tilde{g}+ \cdots$. They may be detectable in colliders in the near future if $\tilde{g}$ is not too heavy \cite{Brignole:1998me}.  Detectable signals may be missing energy, or the production of SM particles as they escape the $\overline{\rm D3}$-branes.

\item In our scenario, the inflatino, the superpartner of the inflaton in the D3-$\overline{\rm D3}$ brane inflationary scenario \cite{Kachru:2003sx}, is not a part of the Goldstino. Since the inflaton (together with the inflatino) is an open string mode stretching between ${\rm D3}$- and $\overline{\rm D3}$-branes (though the attractive force between them comes from the exchange of graviton and RR fields which are closed string modes). Towards the end of inflation, the inflaton dis-appears from the spectrum when the annihilation of brane-anti-branes releases energy to reheat the universe.  Furthermore, the inflation scale ($\sim 10^{14}$--$10^{15}\; {\rm GeV}$~\cite{Kachru:2003sx}) is just a little below the string scale ($M_{\rm S} \sim 10^{16}\;{\rm GeV}$), which is much higher than $\tilde{m}$, so we expect the inflationary scenario to be dominated by $\overline{\rm D3}$-branes in a different throat than the throat we now live in.
This picture is a bit different from that in ref.~\cite{Dudas:2021njv} (see also~\cite{Burgess:2021obw}).

\item The gravitino outside the $\overline{\rm D3}$-branes does pick up a tiny mass coming from the RKU mechanism:
$m_{3/2} = e^{K/2} |W|/M_{\rm Pl}^2 \sim 10^{-30}\, {\rm eV}$~\eqref{gmassRKU}.
Here the Goldstino is a closed string mode, a combination of the superpartners of the K\"ahler, the dilaton and the complex structure modes. This gravitino mass is negligibly small (much smaller than the bound~\eqref{gmassb}) as far as its contribution to dark matter density is concerned.

\end{itemize}

\section{Remarks}
\label{sec:remarks}

In the model for scenario (1), we see how a naturally small $\Lambda$ combined with particle physics phenomenology yields interesting consequences in physics and cosmology that can shed light on a number of high energy physics and cosmological puzzles. Furthermore, it makes specific predictions that can be tested in the near future. 

\begin{itemize}

\item String theory naturally introduces a landscape through dynamical flux compactification, where various vacua (solutions) are located in. Each classically stable solution corresponds to a specific choice of values for the discretized fluxes, which in turn corresponds to a certain type of geometry and compactification.  In the above model, the $\overline{\rm D3}$-branes live in stacks of D7-branes which wrap 4-cycles in the orientifold. 
In this RKU model~\cite{Sumitomo:2013vla,Qiu:2020los}, scanning over a ``dense discretuum" of flux values (in practice, scanning over all the parameters of the model), the resulting properly normalized probability distribution $P(\Lambda) \propto \Lambda^{-1+1/2N}$, where the larger stack of D7-branes has a $SU(N)$ gauge symmetry. This strongly suggests that a small $\Lambda$ can be natural. To assure of this property, the contributions of SUSY breaking and the SSB of the standard model to the vacuum energy must precisely cancel each other. This is realized in the $\overline{\rm D3}$-brane picture with the nilpotent $X$ where the usual gravitino problem is automatically absent. 

\item  In scenario (1), although the $\mu$ term in $W$ contributes neither to $\langle D_HW \rangle$ nor $\langle V \rangle$ via $V_F$, it still contributes to $\langle W \rangle = \langle g(\maths{H})\rangle \simeq \langle \mu H_u H_d \rangle \sim m_{\rm EW}^3$. Since $W$ has mass dimension 3, its contribution to the vacuum energy density goes like $\langle V \rangle \simeq |W|^2/M_{\rm Pl}^2$, or $\mathcal{O} (m_{\rm EW}^6/M_{\rm Pl}^2)$.  In the RKU model, choosing the median of the probability distribution $P(\Lambda)$ to match the observed $\Lambda = \langle V_{\rm RKU} \rangle\simeq 10^{-120} M_{\rm Pl}^4$ implies that $\langle W \rangle = |\mu H_u H_d| \sim |100 \,{\rm GeV}|^3$~\cite{Andriolo:2018dee}, where the complex structure contribution to $W$ is expected to be negligibly small. That is, the electroweak scale ($\sim 100$ GeV)  is intimately tied to the observed value of $\Lambda$.

\item The closed string modes (the K\"ahler moduli, the complex structure moduli and the dilaton) couple to the $\overline{\rm D3}$-branes \cite{Cascales:2003wn,Cribiori:2019hod,Parameswaran:2020ukp}. Since these modes are expected to be very light and rather weakly interacting, one naturally obtain a simple extension of $V_X$~\eqref{Vx},
\[
V_X\to\left|m^2F(s,a) -\kappa \phi^\dagger \phi \right|^2\;, \quad  F(s,a)=1 + \sum_j \left(D_j s_j  + \cdots \right) +\sum_i \left( C_i a_i^2 + \cdots \right)\;,
\]
where the expectation values of the scalars $s_j$ and the pseudo-scalars $a_i$ will roll down to zero (i.e. $F(s_j, a_i) \to 1$) as the universe expands. One with mass $10^{-22}$ eV can roll down and contribute to the dark matter density as the fuzzy dark matter~\cite{Hu:2000ke}. In the axi-Higgs model~\cite{Fung:2021wbz}, $F(a)=1 +Ca^2$ where an axion-like $a$ field with mass $m_a \sim 10^{-30}$--$10^{-29}\,{\rm eV}$ can explain the Lithium puzzle in big bang nucleosynthesis, the Hubble tension as well as the isotropic cosmic birefringence in the cosmic microwave background radiation. This axion may also leave detectable signatures in atomic clock and/or high redshift quasars measurements.
 
\item Towards the end of D3-$\overline{\rm D3}$-brane inflation, cosmic superstrings were produced. Their observational detection will provide a major support for string theory.

\item In scenario (2), although the phenomenology closely tracks that of the MSSM or its extension, the gravitino problem in MSSM is absent simply by transforming the linear SUGRA to a NSUGRA, where the Goldstino $G$ is identified to be an open string mode. Here the brane world scenario in string theory offers a simple resolution to the problem.

\end{itemize}

\begin{acknowledgments}
	We thank Wilfried Buchmuller, Emilian Dudas, Vic Law, Tai Kai Ng, Fernando Quevedo and Timm Wrase for valuable discussions and comments. This work is supported in part by the AOE grant AoE/P-404/18-6 and the CRF grant C6017-20G issued by the Research Grants Council (RGC) of the Government of the Hong Kong SAR China.
\end{acknowledgments}
	
\appendix

\section{D7-Branes}
\label{D7A}

It is known in string theory that, upon dimensional reduction, a stack of $N$ D7-branes wrapping a 4-cycle in the orientifold can lead to an effective 4-dimensional ${\cal{N}}=1$ supersymmetric $SU(N)\times U(1)$ pure gauge theory, which can generate a non-perturbative term in the potential from its gaugino condensate.  
As illustrated in table~\ref{tab:d7}, our scenario can accommodate 3 different stacks of D7-branes, hence 3 gaugino condensates yielding 3 non-perturbative terms.  The RKU scenario requires more than one such non-perturbative term. The RKU model with 3 non-perturbative terms has been studied in ref.~\cite{Andriolo:2019gcb}. If open strings stretching between the $\overline{\rm D3}$-branes and the $U(1)$s of the 3 stacks of D7-branes can yield some quarks and leptons, say the $\overline{5}$s in the SU(5) grand unified theory, it is tantalizing to speculate this is the origin of the 3 families of quarks and leptons.

\begin{table}
\centering
\begin{tabular}{c c c c c c c c c c c }
\toprule
dim & $0$ & $1$ & $2$ & $3$ & $(4$ & $5)$ & $(6$ & $7)$ & $(8$ & $9)$ \\
\midrule
$\overline{\rm D3}$ & $\times$ & $\times$ & $\times$ & $\times$ &  &  &  &  &  &  \\
${\rm D7}_1$ & $\times$ & $\times$ & $\times$ & $\times$ & $\times$ & $\times$ & $\times$ & $\times$ & &  \\
${\rm D7}_2$ & $\times$ & $\times$ & $\times$ & $\times$ & $\times$ & $\times$ &  &  & $\times$ & $\times$  \\
${\rm D7}_3$ & $\times$ & $\times$ & $\times$ & $\times$ &  &  & $\times$ & $\times$ & $\times$ & $\times$ \\
\bottomrule
\end{tabular}
\caption{Table indicates how 3 stacks of D7-branes can appear, each with $SU(N_i)\times U(1)_i$ symmetry, $i=1,2,3$.  Placing the 4-cycles to lie mostly in the dimensions shown, the gaugino condensates can provide 3 non-perturbative terms. $0$-dim refers to the time and others refer to spatial dimensions. The Racetrack K\"ahler Uplift model requires  2  or more such terms.}
\label{tab:d7}
\end{table}

In the presence of the non-perturbative terms, the K\"ahler modulus $T$ is dynamically  stabilized~\cite{Kachru:2003sx,Sumitomo:2013vla}. The value $T+\overline{T}$ that appears in \eqref{Vx} is related to the dimensionless compactification volume $\mathcal{V}$,
\begin{equation}
\mathcal{V}=\left(\frac{M_{\rm Pl}}{M_{\rm S}}\right)^{2}\simeq \left(T+\overline{T}\right)^{3/2} \sim 10^6\;.
\label{eq:ms}
\end{equation}

\section{Following a Massive Photon/Gravitino}	
\label{MGF}

For simplicity, we consider only the $x$-direction of~\eqref{KGA}. Since $A_\mu(x)$ is a real function of spacetime, we make the decomposition 
\begin{equation}
A_\mu(x,t)=a_{\mu}(x) e^{i \omega t} + a^*_{\mu}(x) e^{-i \omega t}\;,
\end{equation}
where $a_{\mu}(x)$ is a complex function of $x$ only. %Here we suppress the $\mu$ index. One should recover it when calculating the field strength. 
Now the equation of motion could be reduced to (where the $\mu$ index is neglected for simplicity)
\begin{equation}\label{eq:a(x)}
-\frac{{\rm d}^2}{{\rm d}x^2} a(x) = \left[\omega^2-V(x)\right]a(x)\;,
\end{equation}
where $V(x>0)=m^2$ and $V(x<0)=0$. It is effectively a time-independent Schr\"odinger equation of a particle in the step potential with unit mass. Therefore, general solutions for two sides are 
\begin{equation}
a(x)=
\begin{cases}
C_+ e^{i \omega x} + C_- e^{-i \omega x} & x<0\\
D_+ e^{i k x} + D_- e^{-i k x} & x>0
\end{cases}\;,
\end{equation}
where $k=\sqrt{\omega^2-m^2}$ could be imaginary. Imposing the boundary condition that both $a(x)$ and $a'(x)$ being continuous at $x=0$~\footnote{If $a'(x)$ is not continuous at the boundary, one would have ambiguity defining the magnetic field $B\sim \partial A$.}, one obtains
\begin{align}
C_+ + C_- & = D_+ + D_- \nonumber\\
\omega \left( C_+ - C_- \right) & = k \left( D_+ - D_- \right)\;.
\label{eq:BC}
\end{align} 
Consider a wave with $\omega> m$ coming from the right $x>0$ and moving towards negative $x$, so $C_+=0$. By solving eq.~\eqref{eq:BC} one would have transmission and reflection probabilities as
\begin{equation}
|T|^2 = \frac{\omega}{k} \left|\frac{C_-}{D_-}\right|^2 = \frac{4 \omega k}{(\omega+k)^2} \;, \quad |R|^2 = \left|\frac{D_+}{D_-}\right|^2 = \left(\frac{\omega-k}{\omega+k}\right)^2\;,
\label{eq:TR}
\end{equation} 
which satisfies $|T|^2+|R|^2=1$. 

If the wave comes from the left side $x<0$ and moves towards increasing $x$, $D_- = 0$. If the energy is enough to penetrate the boundary, one would have a transmission and reflection probability the same as in eq.~\eqref{eq:TR}.
However, if the energy is less than the step potential, that is $\omega<m$, $k=i \sqrt{m^2-\omega^2}$ is imaginary, so the amplitude decays inside $x>0$, $a = D_+e^{-|k|x}=D_+ e^{-\sqrt{m^2-\omega^2} x} $. For incoming energy $\omega=0$, one would have $A_\mu \propto e^{-m x}$, which is the Meissner effect. 

Similarly, this study may be applied to the gravitino equation~\eqref{Gravitinoeq1} where $a_{\mu}$ stands for a spinor in \eqref{eq:a(x)}. For the case of a finite step potential barrier with width $l$ (the thickness of the stack of $\overline{\rm D3}$-branes) in a compactified dimension $L$ (measuring the size of the bottom of the warp throat), one could simply add the periodic boundary condition to the above scattering case by identifying $x \sim x + L$, where $x$ is a direction perpendicular to the 3 large dimensions. 

\begin{figure}
\centering
\includegraphics[scale=1]{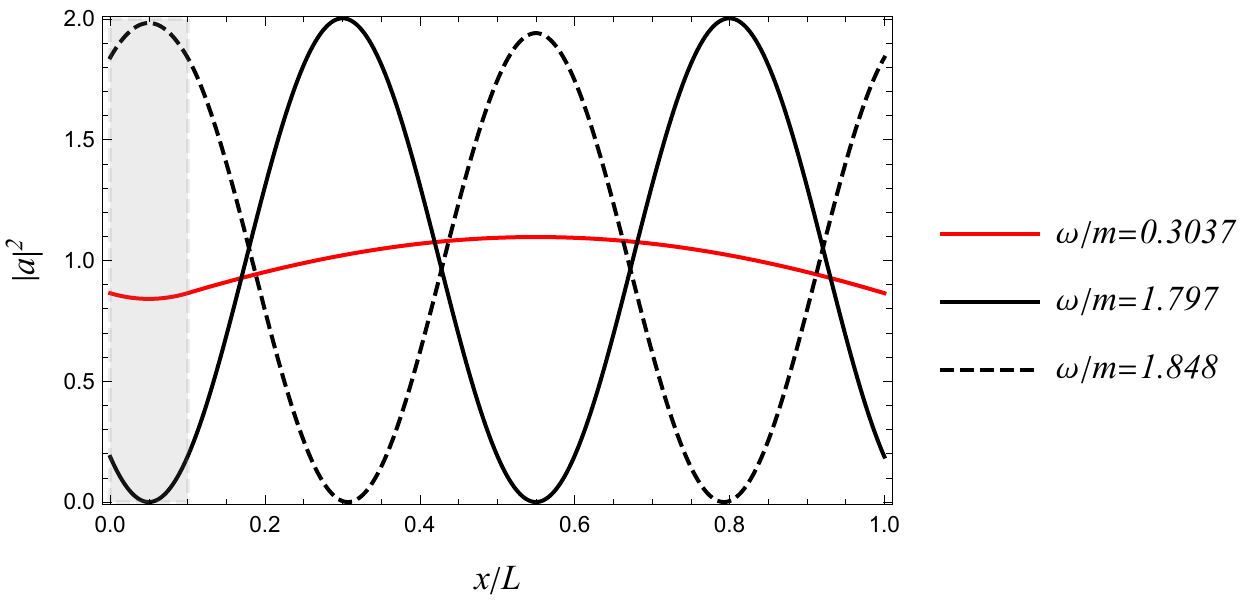}
\caption{An illustration. The probability density distribution of eigen-wavefunctions of 1D time-independent Schr\"odinger equation in a compactified spacetime, $x\sim x+L$. The gray region labels the potential barrier, which is chosen $l/L=0.1$ and $mL=3.5$. The red line labels the only state with energy $\omega<m$.}
\label{fig:1DQM}
\end{figure}

As expected, the energy is quantized. 
Let us illustrate this with the $l/L=0.1$ example here. There is one state with energy smaller than the barrier, $\omega<m$. The probability density of the first three states are shown in figure~\ref{fig:1DQM}. 
Consider the probability of the particle appearing inside the $\overline{\rm D3}$-branes along the $x$-direction as 
\[
P_{\rm inside}=\frac{\int_0^l |a(x)|^2{\rm d} x}{\int_0^L|a(x)|^2{\rm d}x}\;.
\]
The first $10$ states are listed in the table~\ref{tab:eigenstates}. Here $P_{\rm inside}$ is bigger (smaller) than $l/L$ for ``even" (``odd") eigen-states, and it tends to converge around $l/L$ for higher eigen-states, as the mass plays a lesser role.

\begin{table}
\centering
\begin{tabular}{c c l}
\toprule
Eigenstates & Energy ($\omega/m$) & $P_{\rm inside}$\\
\midrule
$0$ &	$0.3037$ &	$0.1407$ \\
$1$ &	$1.797$ &	$0.006320$ \\
$2$ &	$1.848$ &	$0.1934$ \\
$3$ &	$3.594$ &	$0.02388$ \\
$4$ &	$3.615$ &	$0.1801$ \\
$5$ &	$5.390$ &	$0.04883$ \\
$6$ &	$5.400$ &	$0.1544$ \\ 
$7$ &	$7.186$ &	$0.07584$ \\
$8$ &	$7.189$ &	$0.1262$ \\
$9$ &	$8.981$ &	$0.09944$ \\
$10$ &	$8.982$ &	$0.1017$ \\
\bottomrule
\end{tabular}
\caption{Eigenstates in the compactified dimension with $l/L=0.1$ and $mL=3.5$.}
\label{tab:eigenstates}
\end{table}
On average, $P_{\rm inside} \simeq (l/L)$. 
In the warp geometry case, we have $e^{-A(r_b)} \sim r_b/Z \sim m/M_S$, or $r_bM_S \sim mZ$, where $r_b$ measures the size of the bottom of the throat.
Let the thickness of the $\overline{\rm D3}$-branes be its Compton wavelength $1/m$. With average bottom size $L_i \sim r_b$, we have 
\[
P_{\rm inside} \simeq \left(\frac{l}{\bar{L}}\right)^5\simeq \left(\frac{1}{m \bar{L}}\right)^5\sim \frac{1}{\left[(M_S e^{-A(r_b)})(e^{+2A(r_b)}r_b)\right]^5}\sim \frac{e^{-5A(r_b)}}{(mZ)^5}\;,
\]
which is exponentially suppressed for $e^{-A(r_b)} \sim 10^{-14}$ and typical values for $mZ \simeq r_bM_S$. This illustrates that a gravitino prefers to be outside the $\overline{\rm D3}$-branes.
Therefore, for small enough $l/L$, one could safely say that the particle prefers to stay outside the brane. This simple analysis does not incorporate the redshift property and the decay/scattering effect, while section~\ref{sec:mG} gives a qualitative discussion on the warp geometry properties. Clearly a more detailed analysis will be important. Nevertheless, it does capture the basic features underlying the general picture.

\bibliographystyle{JHEP}
\bibliography{ref_gravitino}

\providecommand{\href}[2]{#2}\begingroup\raggedright\begin{thebibliography}{10}

\bibitem{Dudas:2021njv}
E.~Dudas, M.~A.~G. Garcia, Y.~Mambrini, K.~A. Olive, M.~Peloso and S.~Verner,
  \emph{{Slow and Safe Gravitinos}},
  \href{https://doi.org/10.1103/PhysRevD.103.123519}{\emph{Phys. Rev. D}
  {\bfseries 103} (2021) 123519}
  [\href{https://arxiv.org/abs/2104.03749}{{\ttfamily 2104.03749}}].

\bibitem{Pagels:1981ke}
H.~Pagels and J.~R. Primack, \emph{{Supersymmetry, Cosmology and New TeV
  Physics}}, \href{https://doi.org/10.1103/PhysRevLett.48.223}{\emph{Phys. Rev.
  Lett.} {\bfseries 48} (1982) 223}.

\bibitem{Weinberg:1982zq}
S.~Weinberg, \emph{{Cosmological Constraints on the Scale of Supersymmetry
  Breaking}}, \href{https://doi.org/10.1103/PhysRevLett.48.1303}{\emph{Phys.
  Rev. Lett.} {\bfseries 48} (1982) 1303}.

\bibitem{Ellis:1982yb}
J.~R. Ellis, A.~D. Linde and D.~V. Nanopoulos, \emph{{Inflation Can Save the
  Gravitino}}, \href{https://doi.org/10.1016/0370-2693(82)90601-3}{\emph{Phys.
  Lett. B} {\bfseries 118} (1982) 59}.

\bibitem{Khlopov:1984pf}
M.~Y. Khlopov and A.~D. Linde, \emph{{Is It Easy to Save the Gravitino?}},
  \href{https://doi.org/10.1016/0370-2693(84)91656-3}{\emph{Phys. Lett. B}
  {\bfseries 138} (1984) 265}.

\bibitem{Kolb:2021xfn}
E.~W. Kolb, A.~J. Long and E.~McDonough, \emph{{Catastrophic production of slow
  gravitinos}}, \href{https://doi.org/10.1103/PhysRevD.104.075015}{\emph{Phys.
  Rev. D} {\bfseries 104} (2021) 075015}
  [\href{https://arxiv.org/abs/2102.10113}{{\ttfamily 2102.10113}}].

\bibitem{Moroi:1993mb}
T.~Moroi, H.~Murayama and M.~Yamaguchi, \emph{{Cosmological constraints on the
  light stable gravitino}},
  \href{https://doi.org/10.1016/0370-2693(93)91434-O}{\emph{Phys. Lett. B}
  {\bfseries 303} (1993) 289}.

\bibitem{Buchmuller:2019gfy}
W.~Buchmuller, V.~Domcke, H.~Murayama and K.~Schmitz, \emph{{Probing the scale
  of grand unification with gravitational waves}},
  \href{https://doi.org/10.1016/j.physletb.2020.135764}{\emph{Phys. Lett. B}
  {\bfseries 809} (2020) 135764}
  [\href{https://arxiv.org/abs/1912.03695}{{\ttfamily 1912.03695}}].

\bibitem{Ferrara:2014kva}
S.~Ferrara, R.~Kallosh and A.~Linde, \emph{{Cosmology with Nilpotent
  Superfields}}, \href{https://doi.org/10.1007/JHEP10(2014)143}{\emph{JHEP}
  {\bfseries 10} (2014) 143} [\href{https://arxiv.org/abs/1408.4096}{{\ttfamily
  1408.4096}}].

\bibitem{Kallosh:2014wsa}
R.~Kallosh and T.~Wrase, \emph{{Emergence of Spontaneously Broken Supersymmetry
  on an Anti-D3-Brane in KKLT dS Vacua}},
  \href{https://doi.org/10.1007/JHEP12(2014)117}{\emph{JHEP} {\bfseries 12}
  (2014) 117} [\href{https://arxiv.org/abs/1411.1121}{{\ttfamily 1411.1121}}].

\bibitem{Kallosh:2015nia}
R.~Kallosh, F.~Quevedo and A.~M. Uranga, \emph{{String Theory Realizations of
  the Nilpotent Goldstino}},
  \href{https://doi.org/10.1007/JHEP12(2015)039}{\emph{JHEP} {\bfseries 12}
  (2015) 039} [\href{https://arxiv.org/abs/1507.07556}{{\ttfamily
  1507.07556}}].

\bibitem{Bergshoeff:2015jxa}
E.~A. Bergshoeff, K.~Dasgupta, R.~Kallosh, A.~Van~Proeyen and T.~Wrase,
  \emph{{$ \overline{\mathrm{D}3} $ and dS}},
  \href{https://doi.org/10.1007/JHEP05(2015)058}{\emph{JHEP} {\bfseries 05}
  (2015) 058} [\href{https://arxiv.org/abs/1502.07627}{{\ttfamily
  1502.07627}}].

\bibitem{Vercnocke:2016fbt}
B.~Vercnocke and T.~Wrase, \emph{{Constrained superfields from an anti-D3-brane
  in KKLT}}, \href{https://doi.org/10.1007/JHEP08(2016)132}{\emph{JHEP}
  {\bfseries 08} (2016) 132}
  [\href{https://arxiv.org/abs/1605.03961}{{\ttfamily 1605.03961}}].

\bibitem{Kallosh:2016aep}
R.~Kallosh, B.~Vercnocke and T.~Wrase, \emph{{String Theory Origin of
  Constrained Multiplets}},
  \href{https://doi.org/10.1007/JHEP09(2016)063}{\emph{JHEP} {\bfseries 09}
  (2016) 063} [\href{https://arxiv.org/abs/1606.09245}{{\ttfamily
  1606.09245}}].

\bibitem{Sumitomo:2013vla}
Y.~Sumitomo, S.~H.~H. Tye and S.~S.~C. Wong, \emph{{Statistical Distribution of
  the Vacuum Energy Density in Racetrack K\"ahler Uplift Models in String
  Theory}}, \href{https://doi.org/10.1007/JHEP07(2013)052}{\emph{JHEP}
  {\bfseries 07} (2013) 052} [\href{https://arxiv.org/abs/1305.0753}{{\ttfamily
  1305.0753}}].

\bibitem{Li:2020rzo}
S.~Y. Li, Y.-C. Qiu and S.~H.~H. Tye, \emph{{Standard Model from A Supergravity
  Model with a Naturally Small Cosmological Constant}},
  \href{https://doi.org/10.1007/JHEP05(2021)181}{\emph{JHEP} {\bfseries 05}
  (2021) 181} [\href{https://arxiv.org/abs/2010.10089}{{\ttfamily
  2010.10089}}].

\bibitem{Martin:1997ns}
S.~P. Martin, \emph{{A Supersymmetry primer}},  in \emph{{Perspectives on
  supersymmetry. Vol.2}}, G.~L. Kane, ed., vol.~21, pp.~1--153, (2010),
  \href{https://arxiv.org/abs/hep-ph/9709356}{{\ttfamily hep-ph/9709356}},
  \href{https://doi.org/10.1142/9789812839657\_0001}{DOI}.

\bibitem{Kachru:2003aw}
S.~Kachru, R.~Kallosh, A.~D. Linde and S.~P. Trivedi, \emph{{De Sitter vacua in
  string theory}},
  \href{https://doi.org/10.1103/PhysRevD.68.046005}{\emph{Phys. Rev.}
  {\bfseries D68} (2003) 046005}
  [\href{https://arxiv.org/abs/hep-th/0301240}{{\ttfamily hep-th/0301240}}].

\bibitem{Balasubramanian:2004uy}
V.~Balasubramanian and P.~Berglund, \emph{{Stringy corrections to Kahler
  potentials, SUSY breaking, and the cosmological constant problem}},
  \href{https://doi.org/10.1088/1126-6708/2004/11/085}{\emph{JHEP} {\bfseries
  11} (2004) 085} [\href{https://arxiv.org/abs/hep-th/0408054}{{\ttfamily
  hep-th/0408054}}].

\bibitem{Westphal:2006tn}
A.~Westphal, \emph{{de Sitter string vacua from Kahler uplifting}},
  \href{https://doi.org/10.1088/1126-6708/2007/03/102}{\emph{JHEP} {\bfseries
  03} (2007) 102} [\href{https://arxiv.org/abs/hep-th/0611332}{{\ttfamily
  hep-th/0611332}}].

\bibitem{Rummel:2011cd}
M.~Rummel and A.~Westphal, \emph{{A sufficient condition for de Sitter vacua in
  type IIB string theory}},
  \href{https://doi.org/10.1007/JHEP01(2012)020}{\emph{JHEP} {\bfseries 01}
  (2012) 020} [\href{https://arxiv.org/abs/1107.2115}{{\ttfamily 1107.2115}}].

\bibitem{deAlwis:2011dp}
S.~de~Alwis and K.~Givens, \emph{{Physical Vacua in IIB Compactifications with
  a Single Kaehler Modulus}},
  \href{https://doi.org/10.1007/JHEP10(2011)109}{\emph{JHEP} {\bfseries 10}
  (2011) 109} [\href{https://arxiv.org/abs/1106.0759}{{\ttfamily 1106.0759}}].

\bibitem{Rocek:1978nb}
M.~Rocek, \emph{{Linearizing the Volkov-Akulov Model}},
  \href{https://doi.org/10.1103/PhysRevLett.41.451}{\emph{Phys. Rev. Lett.}
  {\bfseries 41} (1978) 451}.

\bibitem{Ivanov:1978mx}
E.~A. Ivanov and A.~A. Kapustnikov, \emph{{General Relationship Between Linear
  and Nonlinear Realizations of Supersymmetry}},
  \href{https://doi.org/10.1088/0305-4470/11/12/005}{\emph{J. Phys. A}
  {\bfseries 11} (1978) 2375}.

\bibitem{Lindstrom:1979kq}
U.~Lindstrom and M.~Rocek, \emph{{CONSTRAINED LOCAL SUPERFIELDS}},
  \href{https://doi.org/10.1103/PhysRevD.19.2300}{\emph{Phys. Rev. D}
  {\bfseries 19} (1979) 2300}.

\bibitem{Komargodski:2009rz}
Z.~Komargodski and N.~Seiberg, \emph{{From Linear SUSY to Constrained
  Superfields}},
  \href{https://doi.org/10.1088/1126-6708/2009/09/066}{\emph{JHEP} {\bfseries
  09} (2009) 066} [\href{https://arxiv.org/abs/0907.2441}{{\ttfamily
  0907.2441}}].

\bibitem{Parameswaran:2020ukp}
S.~Parameswaran and F.~Tonioni, \emph{{Non-supersymmetric String Models from
  Anti-D3-/D7-branes in Strongly Warped Throats}},
  \href{https://arxiv.org/abs/2007.11333}{{\ttfamily 2007.11333}}.

\bibitem{DallAgata:2015zxp}
G.~Dall'Agata and F.~Farakos, \emph{{Constrained superfields in Supergravity}},
  \href{https://doi.org/10.1007/JHEP02(2016)101}{\emph{JHEP} {\bfseries 02}
  (2016) 101} [\href{https://arxiv.org/abs/1512.02158}{{\ttfamily
  1512.02158}}].

\bibitem{Qiu:2020los}
Y.-C. Qiu and S.-H.~H. Tye, \emph{{Linking the Supersymmetric Standard Model to
  the Cosmological Constant}},
  \href{https://arxiv.org/abs/2006.16620}{{\ttfamily 2006.16620}}.

\bibitem{Andriolo:2019gcb}
S.~Andriolo, S.~Y. Li and S.-H.~H. Tye, \emph{{String Landscape and Fermion
  Masses}}, \href{https://doi.org/10.1103/PhysRevD.101.066005}{\emph{Phys. Rev.
  D} {\bfseries 101} (2020) 066005}
  [\href{https://arxiv.org/abs/1902.06608}{{\ttfamily 1902.06608}}].

\bibitem{Dine:1985rz}
M.~Dine, R.~Rohm, N.~Seiberg and E.~Witten, \emph{{Gluino Condensation in
  Superstring Models}},
  \href{https://doi.org/10.1016/0370-2693(85)91354-1}{\emph{Phys. Lett. B}
  {\bfseries 156} (1985) 55}.

\bibitem{Derendinger:1985kk}
J.~Derendinger, L.~E. Ibanez and H.~P. Nilles, \emph{{On the Low-Energy d = 4,
  N=1 Supergravity Theory Extracted from the d = 10, N=1 Superstring}},
  \href{https://doi.org/10.1016/0370-2693(85)91033-0}{\emph{Phys. Lett. B}
  {\bfseries 155} (1985) 65}.

\bibitem{Shifman:1987ia}
M.~A. Shifman and A.~Vainshtein, \emph{{On Gluino Condensation in
  Supersymmetric Gauge Theories. SU(N) and O(N) Groups}},
  \href{https://doi.org/10.1016/0550-3213(88)90680-3}{\emph{Sov. Phys. JETP}
  {\bfseries 66} (1987) 1100}.

\bibitem{Berg:2004ek}
M.~Berg, M.~Haack and B.~Kors, \emph{{Loop corrections to volume moduli and
  inflation in string theory}},
  \href{https://doi.org/10.1103/PhysRevD.71.026005}{\emph{Phys. Rev.}
  {\bfseries D71} (2005) 026005}
  [\href{https://arxiv.org/abs/hep-th/0404087}{{\ttfamily hep-th/0404087}}].

\bibitem{Maltoni:2015twa}
F.~Maltoni, A.~Martini, K.~Mawatari and B.~Oexl, \emph{{Signals of a superlight
  gravitino at the LHC}},
  \href{https://doi.org/10.1007/JHEP04(2015)021}{\emph{JHEP} {\bfseries 04}
  (2015) 021} [\href{https://arxiv.org/abs/1502.01637}{{\ttfamily
  1502.01637}}].

\bibitem{Hebecker:2019csg}
A.~Hebecker, T.~Skrzypek and M.~Wittner, \emph{{The $F$-term Problem and other
  Challenges of Stringy Quintessence}},
  \href{https://doi.org/10.1007/JHEP11(2019)134}{\emph{JHEP} {\bfseries 11}
  (2019) 134} [\href{https://arxiv.org/abs/1909.08625}{{\ttfamily
  1909.08625}}].

\bibitem{Dudas:2015eha}
E.~Dudas, S.~Ferrara, A.~Kehagias and A.~Sagnotti, \emph{{Properties of
  Nilpotent Supergravity}},
  \href{https://doi.org/10.1007/JHEP09(2015)217}{\emph{JHEP} {\bfseries 09}
  (2015) 217} [\href{https://arxiv.org/abs/1507.07842}{{\ttfamily
  1507.07842}}].

\bibitem{Klebanov:2000hb}
I.~R. Klebanov and M.~J. Strassler, \emph{{Supergravity and a confining gauge
  theory: Duality cascades and chi SB resolution of naked singularities}},
  \href{https://doi.org/10.1088/1126-6708/2000/08/052}{\emph{JHEP} {\bfseries
  08} (2000) 052} [\href{https://arxiv.org/abs/hep-th/0007191}{{\ttfamily
  hep-th/0007191}}].

\bibitem{Brignole:1998me}
A.~Brignole, F.~Feruglio, M.~L. Mangano and F.~Zwirner, \emph{{Signals of a
  superlight gravitino at hadron colliders when the other superparticles are
  heavy}}, \href{https://doi.org/10.1016/S0550-3213(98)00254-5}{\emph{Nucl.
  Phys. B} {\bfseries 526} (1998) 136}
  [\href{https://arxiv.org/abs/hep-ph/9801329}{{\ttfamily hep-ph/9801329}}].

\bibitem{Kachru:2003sx}
S.~Kachru, R.~Kallosh, A.~D. Linde, J.~M. Maldacena, L.~P. McAllister and S.~P.
  Trivedi, \emph{{Towards inflation in string theory}},
  \href{https://doi.org/10.1088/1475-7516/2003/10/013}{\emph{JCAP} {\bfseries
  0310} (2003) 013} [\href{https://arxiv.org/abs/hep-th/0308055}{{\ttfamily
  hep-th/0308055}}].

\bibitem{Burgess:2021obw}
C.~P. Burgess, D.~Dineen and F.~Quevedo, \emph{{Yoga Dark Energy: natural
  relaxation and other dark implications of a supersymmetric gravity sector}},
  \href{https://doi.org/10.1088/1475-7516/2022/03/064}{\emph{JCAP} {\bfseries
  03} (2022) 064} [\href{https://arxiv.org/abs/2111.07286}{{\ttfamily
  2111.07286}}].

\bibitem{Andriolo:2018dee}
S.~Andriolo, S.~Y. Li and S.~H.~H. Tye, \emph{{The Cosmological Constant and
  the Electroweak Scale}},
  \href{https://doi.org/10.1007/JHEP10(2019)212}{\emph{JHEP} {\bfseries 10}
  (2019) 212} [\href{https://arxiv.org/abs/1812.04873}{{\ttfamily
  1812.04873}}].

\bibitem{Cascales:2003wn}
J.~F.~G. Cascales, M.~P. Garcia~del Moral, F.~Quevedo and A.~M. Uranga,
  \emph{{Realistic D-brane models on warped throats: Fluxes, hierarchies and
  moduli stabilization}},
  \href{https://doi.org/10.1088/1126-6708/2004/02/031}{\emph{JHEP} {\bfseries
  02} (2004) 031} [\href{https://arxiv.org/abs/hep-th/0312051}{{\ttfamily
  hep-th/0312051}}].

\bibitem{Cribiori:2019hod}
N.~Cribiori, C.~Roupec, T.~Wrase and Y.~Yamada, \emph{{Supersymmetric
  anti-D3-brane action in the Kachru-Kallosh-Linde-Trivedi setup}},
  \href{https://doi.org/10.1103/PhysRevD.100.066001}{\emph{Phys. Rev. D}
  {\bfseries 100} (2019) 066001}
  [\href{https://arxiv.org/abs/1906.07727}{{\ttfamily 1906.07727}}].

\bibitem{Hu:2000ke}
W.~Hu, R.~Barkana and A.~Gruzinov, \emph{{Cold and fuzzy dark matter}},
  \href{https://doi.org/10.1103/PhysRevLett.85.1158}{\emph{Phys. Rev. Lett.}
  {\bfseries 85} (2000) 1158}
  [\href{https://arxiv.org/abs/astro-ph/0003365}{{\ttfamily
  astro-ph/0003365}}].

\bibitem{Fung:2021wbz}
L.~W.~H. Fung, L.~Li, T.~Liu, H.~N. Luu, Y.-C. Qiu and S.~H.~H. Tye,
  \emph{{Axi-Higgs cosmology}},
  \href{https://doi.org/10.1088/1475-7516/2021/08/057}{\emph{JCAP} {\bfseries
  08} (2021) 057} [\href{https://arxiv.org/abs/2102.11257}{{\ttfamily
  2102.11257}}].

\end{thebibliography}\endgroup

\end{document}